\documentclass[aps,amsmath,notitlepage,twocolumn,amssymb,prb,letterpaper,longbibliography]{revtex4-1}
\usepackage[pdftex]{graphicx} 
\usepackage{epstopdf}
\usepackage{verbatim}
\usepackage{color}
\usepackage{subfigure}
\usepackage{tabularx}
\usepackage{amsfonts}
\usepackage{wasysym}
\usepackage{bm}
\usepackage{multirow}

\usepackage[colorlinks=true,citecolor=red,urlcolor=blue]{hyperref}

\definecolor{darkred}{rgb}{0.847059, 0.141176, 0.164706}
\definecolor{darkgreen}{rgb}{0,0.4,0}
\definecolor{darkblue}{rgb}{0.254902, 0.411765, 0.882353}

\newcommand{\Rmnum}[1]{\uppercase\expandafter{\romannumeral #1\relax}}

\begin{document}

\title{Possible gapless spin liquid in a rare-earth kagom\'{e} lattice magnets Tm$_{3}$Sb$_{3}$Zn$_{2}$O$_{14}$}

\author{Zhao-Feng Ding$^{1}$}
\author{Yan-Xing Yang$^{1}$}
\author{Jian Zhang$^{1}$}
\author{Cheng Tan$^{1}$}
\author{Zi-Hao Zhu$^{1}$}
\author{Gang Chen$^{1,2,3}$}
\email{gangchen.physics@gmail.com}
\author{Lei Shu$^{1,3}$}
\email{leishu@fudan.edu.cn}
\affiliation{$^{1}$State Key Laboratory of Surface Physics,
Department of Physics, Fudan University, Shanghai 200433, China}
\affiliation{$^{2}$Center for Field Theory and Particle Physics,
Fudan University, Shanghai 200433, China}
\affiliation{$^{3}$Collaborative Innovation Center of Advanced Microstructures,
Nanjing University, Nanjing 210093, China}

\date{\today}

\begin{abstract}
We report the thermodynamic and muon spin relaxation ($\mu$SR)
evidences for a possible gapless spin liquid in Tm$_{3}$Sb$_{3}$Zn$_{2}$O$_{14}$, with the rare-earth
ions Tm$^{3+}$ forming a two-dimensional kagom\'{e} lattice.
We extract the magnetic specific heat of Tm$_{3}$Sb$_{3}$Zn$_{2}$O$_{14}$
by subtracting the phonon contribution of the non-magnetic isostructural
material La$_{3}$Sb$_{3}$Zn$_{2}$O$_{14}$ and obtain a clear linear-$T$
temperature dependence of magnetic specific heat at low temperatures.
No long-range magnetic order was observed down to 0.35 K in the heat capacity measurements, and
no signature of spin freezing down to 50 mK was observed in A.C. susceptibility measurements.
The absence of magnetic order is further confirmed by the $\mu$SR
measurements down to 20 mK. We find that the spin-lattice
relaxation time remains constant down to the lowest temperature.
We point out that the physics in Tm$_{3}$Sb$_{3}$Zn$_{2}$O$_{14}$
is fundamentally different from the Cu-based herbertsmithite and
propose spin liquid ground states with non-Kramers doublets on the
kagom\'{e} lattice to account for the experimental results. However, we can not rule out that these exotic properties are induced by the Tm/Zn site-mixing disorder in Tm$_{3}$Sb$_{3}$Zn$_{2}$O$_{14}$.
\end{abstract}

\maketitle

\section{Introduction}

Quantum spin liquid (QSL) is an exotic quantum state of matter in which the
spins are highly entangled and remain disordered even down to zero temperature.
The search of QSLs has attracted a significant attention partly due to its
potential application on quantum information and the possible relevance to
high superconductivity~\cite{Anderson87,Lee08,Balents10,Zhou17}.
As a new quantum phase of matter beyond the traditional Landau's symmetry
breaking paradigm, QSL has its own value. Instead of associating with
certain order parameters for conventional symmetry breaking states,
QSL is often characterized by the excitations with fractionalized spin
quantum number and the emergent gauge structure. Precisely due to the
absence of magnetic order, the experimental identification of QSLs requires
more scrutiny than conventional orders. In the last decade or so, an increasing
number of new materials have been proposed to be QSL candidates. These spin
liquid candidates often have the lattice structures with geometric frustration
such as triangle lattice ($\kappa$-[ET]$_2$Cu$_2$(CN)$_3$~\cite{Shimizu03,Shimizu06,Ohira07,Yamashita08,Pratt11,Nakajima12,Isono16},
EtMe$_3$Sb[Pd(dmit)$_2$]$_2$~\cite{Itou07,Itou08,Itou10,Shaginyan13}, and YbMgGaO$_4$~\cite{srep,YueshengPRL,YaodongPRB,Shen16,Li16,Martin17,Shiyan2016,PhysRevB.96.054445,PhysRevB.94.174424,
PhysRevLett.119.157201,PhysRevLett.120.037204,PhysRevB.96.075105,Yaodong201608,yuesheng1702,Toth1705,JunZhaounpub,PeterArmitagearXiv,Ma18}),
kagom\'{e} lattice (ZnCu$_3$(OH)$_6$Cl$_2$~\cite{Helton07,YLeeNature2012,FuNMRScience2015}),
hyperkagom\'{e} lattice (Na$_4$Ir$_3$O$_8$~\cite{Okamoto07,PhysRevB.78.094403,PhysRevLett.101.197201,
PhysRevLett.112.087401,Chen2013,PhysRevLett.113.247601}),
and pyrochlore lattice (Yb$_2$Ti$_2$O$_7$~\cite{Yaouanc11,Ross11}, Pr$_2$Zr$_2$O$_7$~\cite{Matsuhira09},
Pr$_2$Ir$_2$O$_7$~\cite{Nakatsuji06,Machida09,Tokiwa14,MacLaughlin15,Chen16,Cheng17,Yao17}, Tb$_2$Ti$_2$O$_7$~\cite{Gingras00,Hiroshi12}, et al.),
and our work here is about a rare-earth kagom\'{e} lattice magnet.

Recently, a new family of pyrochlore derivatives
\emph{RE}$_{3}$Sb$_{3}$\emph{M}$_{2}$O$_{14}$ (\emph{RE}
= La, Pr, Nd, Gd, Tb, Dy, Ho, Er, and Yb; \emph{M} = Mg, Zn)
with a kagom\'{e} lattice structure was discovered, and their
crystal structure, thermal transport, and magnetic properties
were studied~\cite{Sanders16,Dun16,Scheie16,Paddison16,Dun17,Dun18}.
In these materials, the rare-earth ions form two-dimensional
kagom\'{e} lattice layers that are separated by non-magnetic
Zn/Mg layer. Several magnetic phases including the emergent
kagom\'{e} Ising order~\cite{Paddison16}, non-magnetic singlet
state~\cite{Dun17}, spin glass~\cite{Dun17}, and even spin
liquid~\cite{Dun17} were proposed.

Despite the previous effort, one compound in this material family
has not been carefully explored. Here, we study the magnetic properties of
the compounds Tm$_{3}$Sb$_{3}$Zn$_{2}$O$_{14}$ in this family
as well as a non-magnetic reference compound
La$_{3}$Sb$_{3}$Zn$_{2}$O$_{14}$. The heat capacity
measurements did not find any evidence of
magnetic order for Tm$_{3}$Sb$_{3}$Zn$_{2}$O$_{14}$ down to
${\sim 0.35}$ K despite an antiferromagnetic Curie-Weiss temperature
${\Theta_{\text{CW}} = -18.6}$ K. A linear-$T$ heat capacity
is obtained at low temperatures, indicating a constant density
of states and gapless excitations at low energies. No spin freezing was
 observed in A.C. magnetic susceptibility measurements.
The absence of the magnetic order is further confirmed by
the muon spin relaxation ($\mu$SR) measurements.
The constant spin lattice relaxation rate at low energies
seems to be consistent with the large and constant
density of states at low energies in this system.
We discuss the microscopic origin of the Tm$^{3+}$ local moments,
propose the candidate spin liquid ground states with either
spinon Fermi surface or spinon quadratic node
and further experiments for Tm$_{3}$Sb$_{3}$Zn$_{2}$O$_{14}$. Inspired by a recent theoretical study of the effect of quenched bond disorder on spin-1/2 quantum magnets~\cite{Kimchi18}, the possibility of Tm/Zn site-mixing disorder induced localized two-level scenario can not be excluded.

\begin{figure}[htp]
  \begin{center}
  \includegraphics[width=8.5cm]{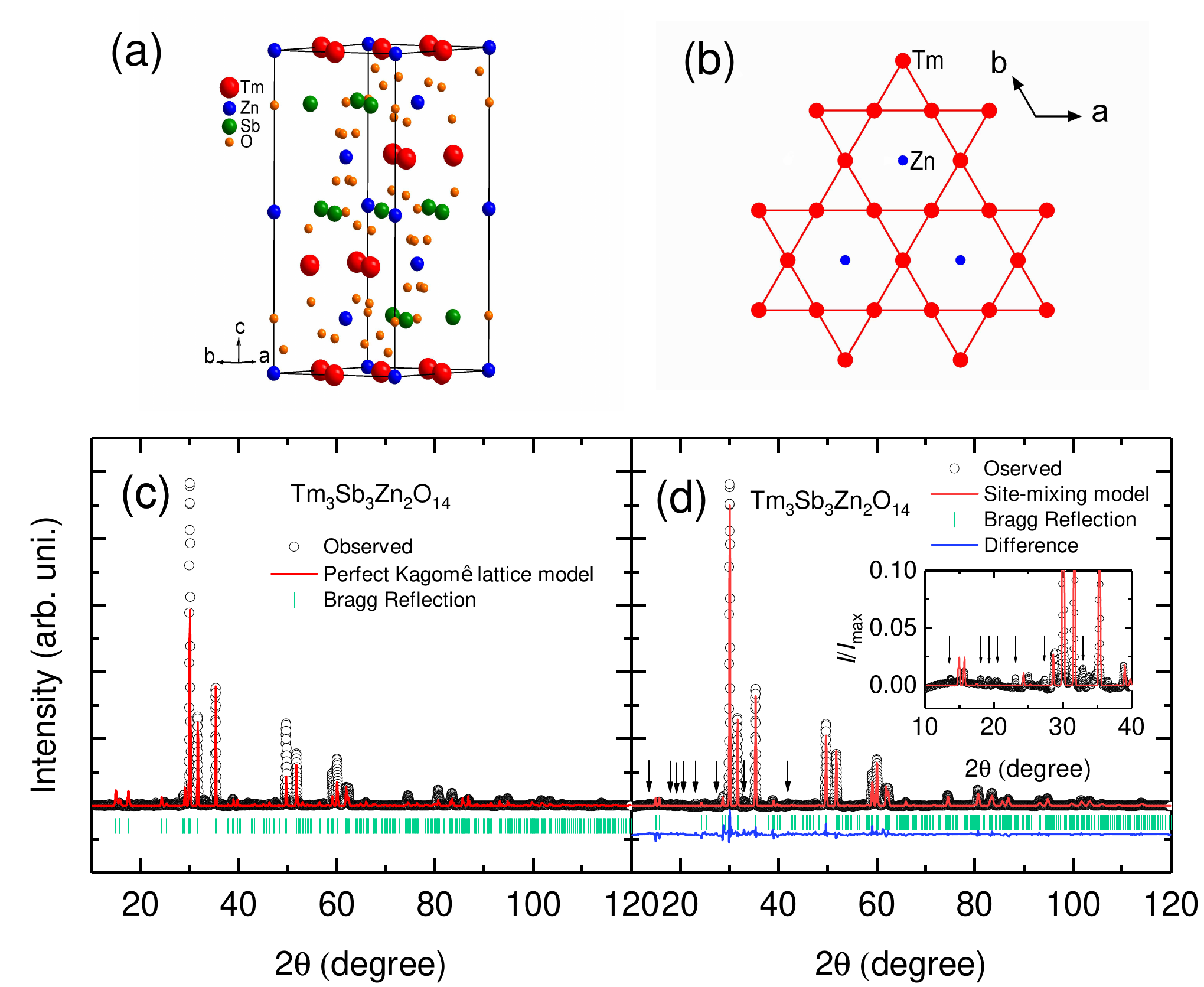}
  \caption{(a) Crystal structure of Tm$_3$Sb$_3$Zn$_2$O$_{14}$ in one unit cell.
  (b) The kagom\'{e} lattice formed by the Tm$^{3+}$ ions in the ab plane. The position of in-plane Zn$^{2+}$ was also shown. (c) Rietveld refinement of powder XRD pattern for Tm$_3$Sb$_3$Zn$_2$O$_{14}$ using the perfect kagom\'{e} lattice model, which cannot describe the data well. The green bars
  indicates the Bragg reflections.
  (d) Rietveld refinement of powder XRD pattern for Tm$_3$Sb$_3$Zn$_2$O$_{14}$ using site-mixing model. The black circles, red line, and blue line are the experimental data, the calculated patterns based on the Tm/Zn site-mixing model and the differences, respectively. The green bars
  indicates the Bragg reflections. Inset is the enlargement of the low degree part. The arrows indicate the tiny peaks that
  can not be described by Tm/Zn site-mixing model.}
  \label{fig1}
  \end{center}
\end{figure}

\section{experimental details}

The polycrystalline \emph{RE}$_{3}$Sb$_{3}$Zn$_{2}$O$_{14}$
(\emph{RE} = La,Tm) samples were synthesized by solid state
reaction method as reported before~\cite{Sanders16}.
The sample structure was confirmed by powder X-ray diffraction
(XRD) measurements (Bruker D8 Advance, ${\lambda = 1.5418}$ \AA) at room
temperature. Rietveld crystal structure refinements of the XRD data were
performed by using GSAS program~\cite{GSAS} and EXPGUI~\cite{EXPGUI}.

Magnetization was measured in a superconducting quantum interference
device magnetometer (Quantum Design Magnetic Property Measurement System) down to 2 K.
A.C. susceptibility was measured in a Quantum Design Physical Property
Measurement System (PPMS) equipped with A.C. Measurement System
for the Dilution Refrigerator option in a temperature range of 0.05 - 4 K.
Specific heat down to 0.35 K was measured by the adiabatic relaxation method on
a Dynacool-PPMS platform equipped with Helium-3 option.

Zero field muon spin relaxation (ZF-$\mu$SR) experiments down to
20 mK were carried out by using the continuous beam line on the DR
spectrometer at TRIUMF, Vancouver, Canada. During the $\mu$SR experiment,
the powder sample was mounted on a silver sample holder using GE-varnish.
Before the experiments, a weak transverse field run was carried out at 6 K to
determine the value of $\alpha$, which is the detector efficiency related
to the experimental setup. ZF-$\mu$SR experiments were then carried out after zeroing the magnetic field
by measuring the standard silver sample with method as introduced
before~\cite{Morris03}.

\section{RESULTS}

\subsection{Crystal Structure}

\begin{table*}[t]
\caption{Rietveld fitting results using both perfect kagom\'{e} lattice model and site-mixing model.}\label{table-A}
\begin{ruledtabular}
\begin{tabular}{lcccccccc}
Atom & Site & $x$ & $y$ & $z$ & Occ.& $a,b$ (\AA) &$c$ (\AA) &$\chi^{2}$\\
\hline
Perfect kagom\'{e} lattice model&&&&&&&&\\
\hline
 Tm &  9 & 0.5        & 0         & 0         & 1& \multirow{7}*{7.3446(2)}&\multirow{7}*{16.9930(6)}&\multirow{7}*{1547}\\
 Zn1&  3 & 0          & 0         & 0         & 1&&&\\
 Zn2& 18 & 0          & 0         & 0.5       & 1&&&\\
 Sb &  9 & 0.5        & 0         & 0.5       & 1&&&\\
 O1 &  6 & 0          & 0         & 0.326(6)  & 1&&&\\
 O2 & 18 & 0.616(4)   &-0.616(4)  & 0.227(3)  & 1&&&\\
 O3 & 18 & 0.196(5)   &-0.196(5)  &-0.038(4)  & 1&&&\\
\hline
Site-mixing model&&&&&&&&\\
\hline
 Tm &  9 & 0.5        & 0         & 0         & 0.827(6)&\multirow{9}*{7.3568(2)}&\multirow{9}*{17.0196(7)}&\multirow{9}*{380}\\
 Zn(disorder) &  9 & 0.5   & 0    & 0         & 0.173(6)&&&\\
 Zn1&  3 & 0          & 0         & 0         & 0.825(6)&&&\\
 Tm(disorder) &  3 & 0& 0         & 0         & 0.175(6)&&&\\
 Zn2& 18 & 0          & 0         & 0.5       & 1&&&\\
 Sb &  9 & 0.5        & 0         & 0.5       & 1&&&\\
 O1 &  6 & 0          & 0         & 0.326(3)  & 1&&&\\
 O2 & 18 & 0.616(2)   &-0.616(2)  & 0.234(1)  & 1&&&\\
 O3 & 18 & 0.147(2)   &-0.147(2)  &-0.052(1)  & 1&&&\\
\end{tabular}
\end{ruledtabular}
\end{table*}

Our sample quality was examined by the XRD and magnetic susceptibility
measurements. The crystal structure of Tm$_{3}$Sb$_{3}$Zn$_{2}$O$_{14}$ is shown in
 Fig.~\ref{fig1}(a). Fig.~\ref{fig1}(b) shows the kagom\'{e} layer formed by Tm$^{3+}$ ions.
 The typical XRD spectrum of Tm$_{3}$Sb$_{3}$Zn$_{2}$O$_{14}$
is depicted in Fig.~\ref{fig1}(c) and (d). From Fig.~\ref{fig1}(c),
we notice that the low degree peaks below 30 degree of Tm$_{3}$Sb$_{3}$Zn$_{2}$O$_{14}$ are absent. This is inconsistent with the perfect kagom\'{e} lattice model.
Similar behavior was also observed in \emph{RE}$_{3}$Sb$_{3}$Zn$_{2}$O$_{14}$
(\emph{RE} is rare earth element with ion radius smaller than Dy) and was explained
by the \emph{RE}/Zn site-mixing disorder in recent works~\cite{Dun17,Dun18}. The XRD pattern of Tm$_{3}$Sb$_{3}$Zn$_{2}$O$_{14}$ is fitted by using the site-mixing model. Table~\ref{table-A} lists the fitting parameters using both perfect kagom\'{e} lattice model without disorder (fitted curve in Fig.~\ref{fig1} (c)) and site-mixing model (fitted curve in Fig.~\ref{fig1} (d)). It can be seen that the site-mixing model describes the data better, and it suggests that only $\sim$82.7$\%$ of the Tm ions stay on their original position and $\sim$82.5$\%$ Zn ions occupy the in-plane Zn sites. There are site-mixting between Tm and in-plane Zn ions. As shown in the main graph of Fig.~\ref{fig1}(d), the site-mixing model can describe the main feature of the XRD pattern, despite of several additional tiny peaks indicated by arrows in inset of Fig.~\ref{fig1} (d).
Similar results were also observed in previous work, and it was suggested that the XRD pattern may be
more accurately described by a different model~\cite{Dun17}. Since better model has not been found, we take the site-mixing of Tm/Zn as the main source of disorder in Tm$_{3}$Sb$_{3}$Zn$_{2}$O$_{14}$. The possible disorder effect will be discussed in the discussion section.

The distorted Tm kagom\'{e} layers are separated by the non-magnetic Zn, Sb and O atoms.
The interlayer separation of the Tm kagom\'{e} layers is 5.67~\AA. Since the Tm layers
are not stacked right on top of each other, the nearest Tm-Tm bond
between two neighboring layers is actually larger than the interlayer
separation and is 6.06~\AA. Within the kagom\'{e} layers,
the nearest Tm-Tm bond is 3.68~\AA\, and is much smaller than
the interlayer Tm-Tm bonds. This justifies the quasi-two-dimensional
nature of Tm$_{3}$Sb$_{3}$Zn$_{2}$O$_{14}$.

\subsection{Magnetization}

\begin{figure}[t]
\begin{center}
\includegraphics[width=8.5cm]{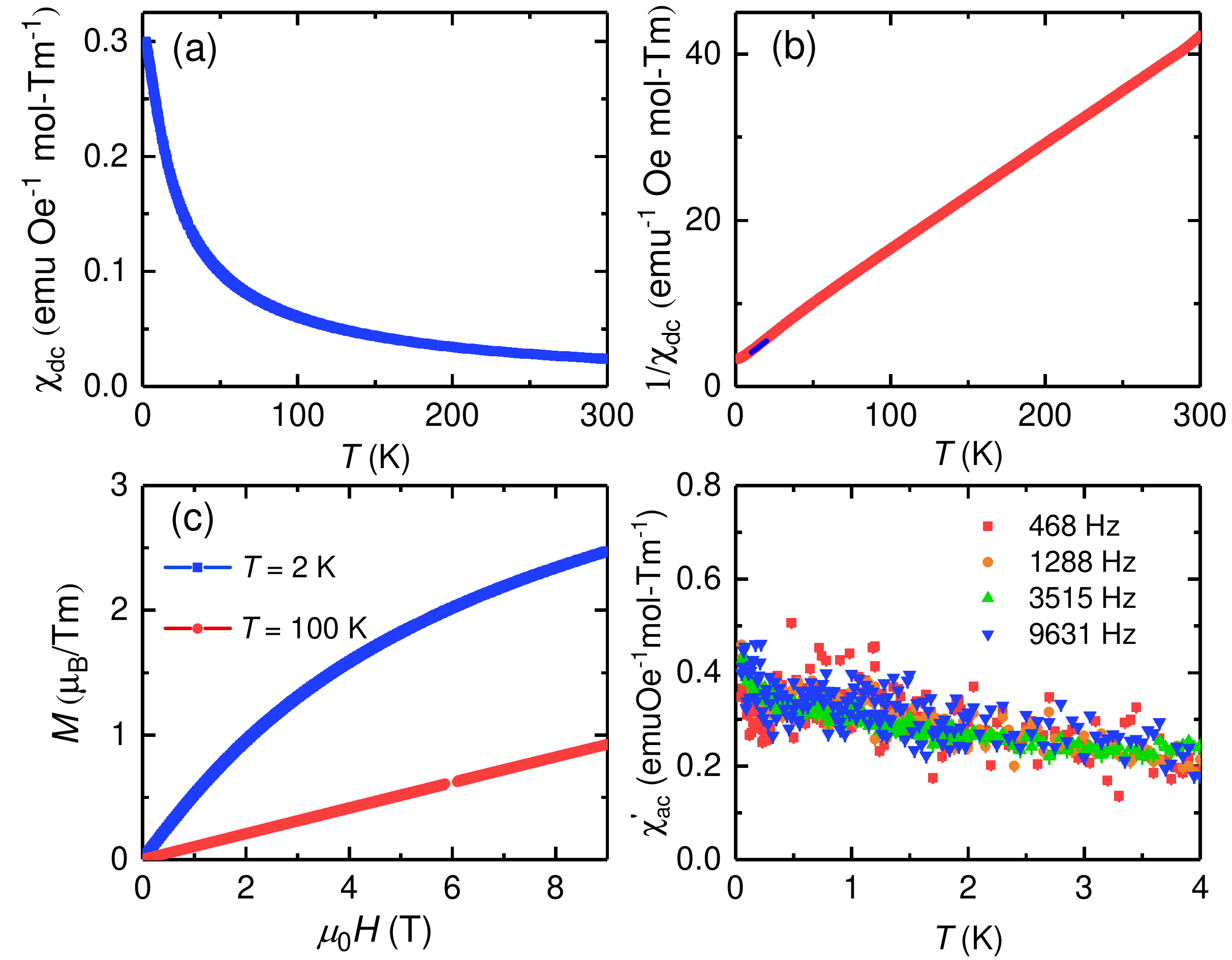}
\caption{Temperature dependence of magnetic susceptibility (a)
and its inverse (b) of Tm$_{3}$Sb$_{3}$Zn$_{2}$O$_{14}$ under an applied magnetic field of $\mu_0$H = 0.01 T. Blue line in (b) is the Curie-Weiss law fitting curve of
the data between 10 and 20 K. (c) Isothermal magnetization up to 9 T at 2 K
(blue squares) and 100 K (red circles). (d) Temperature dependence of real
part of A.C. susceptibility with different driving frequencies. The low frequency results were shifted upwards for comparison. No spin freezing
behaviour was observed down to 50 mK.}
\label{fig2}
\end{center}
\end{figure}

The magnetic susceptibility of Tm$_{3}$Sb$_{3}$Zn$_{2}$O$_{14}$
was measured under an external field of $\mu_0$H = 0.01 T down
to 2 K and is depicted in Fig.~\ref{fig2}(a). As shown in Fig.~\ref{fig2}(b), the Curie-Weiss
law behaviour is observed in high temperature range. The
Curie-Weiss law fitting in the range of 10 - 20 K (the most Curie-Weiss like region, just above the Schottky anomaly observed in specific heat) yields a
Curie-Weiss temperature ${\Theta_{\text{CW}} = -18.6}$ K
and an effective magnetic moment $\mu_{\text{eff}}$ = 7.50 $\mu_B$.
The effective moment here arises from the combination
of crystal electric field and the atomic spin orbit coupling.
The Tm$^{3+}$ ion has an electron configuration $4f^{12}$,
and atomic spin orbit coupling gives a total angular momentum
$J = 6$. The crystal electric field splits the 13-fold degeneracy of
the total moment $J$, and the ground state doublet is a non-Kramers
doublet and is regarded as an effective spin-1/2 moment.
In principle, the crystal symmetry of the kagom\'{e} lattice
forbids any two-fold degeneracy for the integer spin moments and
would necessarily create an energy splitting between the two states
of the non-Kramers doublet. However, if the energy splitting between
these two states is much smaller than the crystal field energy gap
that separates these two states from other excited ones,
then one can still think the low-temperature magnetic properties
of the system are from these non-Kramers doublets.
We will discuss this in details in the later part of this article.
The negative $\Theta_{\text{CW}}$ reflects the antiferromagnetic
nature of the exchange interactions between the Tm$^{3+}$ local
moments in Tm$_{3}$Sb$_{3}$Zn$_{2}$O$_{14}$. The isothermal magnetization
up to 9 T at 2 K and 100 K are shown in Fig.~\ref{fig2}(c).
At 2 K, the M-H curve starts to deviate from the linear
response regime around 2 T, which is consistent with the
relatively low energy scale of the interaction between the
(low-lying) non-Kramers doublets of the Tm$^{3+}$ ions.
In contrast, at 100 K, the excited crystal field states of the
Tm$^{3+}$ ions would be thermally excited and contribute to the
magnetization. This prevents the magetization from saturation,
and the M-H curve remains linear within our field range. As shown
in Fig.~\ref{fig2} (d), A.C. susceptibility of Tm$_{3}$Sb$_{3}$Zn$_{2}$O$_{14}$
under different driving frequencies was measured down to 50 mK .
The Curie-Weiss like behaviour instead of the down turn feature of real part of
A.C. susceptibility at low temperatures indicates the absence of spin
glass behaviour. Additionally, no signature of transition or separation
between zero-field cooling and field cooling procedures (not shown here)
was observed, eliminating the possibility of spin glass state.

\subsection{Specific Heat}

\begin{figure}[t]
  \begin{center}
  \includegraphics[width=8.5cm]{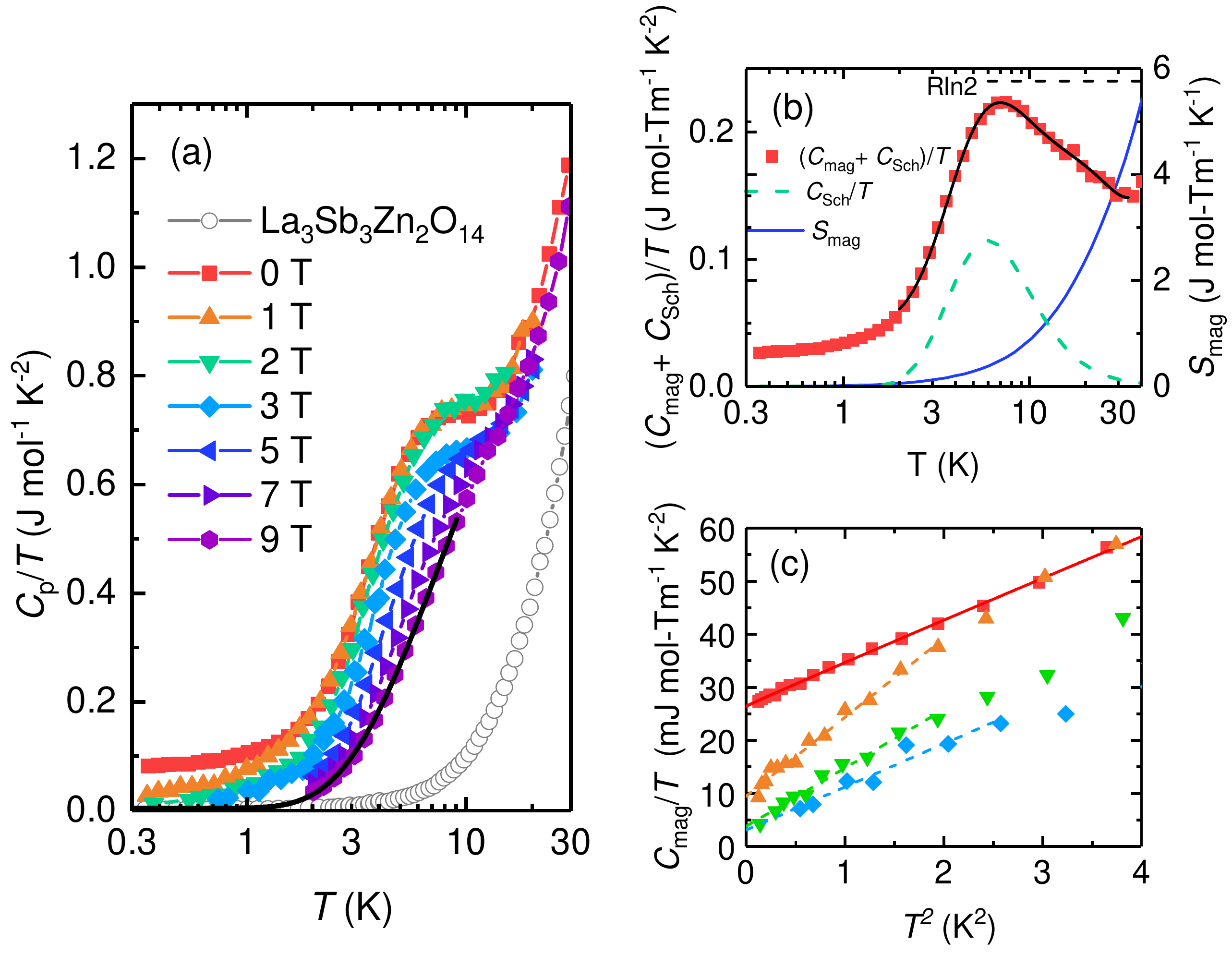}
  \caption{(a) Specific heat of Tm$_{3}$Sb$_{3}$Zn$_{2}$O$_{14}$ at different magnetic fields up to 9 T. A Schottky type kink is observed around 9 K at zero field and is gradually
  suppressed by applying magnetic fields. The solid black line shows the exponential fitting of the 9 T data as
  described in the main text. The specific heat of non-magnetic La$_{3}$Sb$_{3}$Zn$_{2}$O$_{14}$ is also shown for comparison. (b) The phonon contribution subtracted specific heat of Tm$_{3}$Sb$_{3}$Zn$_{2}$O$_{14}$ at zero field. The green dashed curve indicates the fitted Schottky contribution. The blue line is the magnetic entropy change obtained
  by integrating the magnetic specific heat from the lowest measured temperature. The horizontal dashed line indicates
  the entropy for spin-1/2 moments R$\ln 2$.
  (c) Low temperature specific heat coefficient ($C_p/T$) per mol Tm$^{3+}$ as a function of $T^{2}$.
  The red line is a linear fit of the zero field data, indicating a finite interception $\gamma (0) = 26.6(1)$
  mJ mol-Tm$^{-1}$ K$^{-2}$. $\gamma (0)$ is gradually suppressed by magnetic fields
  and finally reaches a finite value around 3 T. The dashed lines are guide to the eye.}
  \label{fig3}
  \end{center}
\end{figure}

The magnetic susceptibility and magnetization measurements provide the
basic information about the magnetic moments and the interaction energy
scale in Tm$_{3}$Sb$_{3}$Zn$_{2}$O$_{14}$. To reveal the low-energy
and low-temperature properties, we measured the specific heat of
Tm$_{3}$Sb$_{3}$Zn$_{2}$O$_{14}$ at different applied magnetic fields.
Thanks to the non-magnetic isostructural material
La$_{3}$Sb$_{3}$Zn$_{2}$O$_{14}$, we are able to obtain
the phonon contribution to the specific heat. In Fig.~\ref{fig3},
we depict the specific heat data of Tm$_{3}$Sb$_{3}$Zn$_{2}$O$_{14}$
and La$_{3}$Sb$_{3}$Zn$_{2}$O$_{14}$. At zero magnetic field, a broad hump
was observed for Tm$_{3}$Sb$_{3}$Zn$_{2}$O$_{14}$ around 9 K,
which can be fitted by adding a Schottky anomaly term (green dashed curve in Fig.~\ref{fig3}(b)),
with the gap value ${\Delta = 19.4(2)}$ K. The Schottky anomaly should come
from the effect of disorder since it was gradually suppressed by the applied
magnetic field. No sharp anomaly for magnetic transitions was observed
down to the lowest measured temperature 0.35 K, indicating the
absence of long range magnetic orders in Tm$_{3}$Sb$_{3}$Zn$_{2}$O$_{14}$.
As is expected, magnetic field does not show any influence on the
specific heat of La$_{3}$Sb$_{3}$Zn$_{2}$O$_{14}$. Moreover, from
the comparison in Fig.~\ref{fig3} (a), the phonon contribution to
the specific heat of Tm$_{3}$Sb$_{3}$Zn$_{2}$O$_{14}$ is almost
negligible below $\sim$4 K.

The low-temperature specific heat coefficients $C_{p}/T$ of Tm$_{3}$Sb$_{3}$Zn$_{2}$O$_{14}$
at different applied magnetic fields are further displayed in Fig.~\ref{fig3} (c).
By extrapolating the zero field $C_{p}/T$ to zero temperature,
a finite interception $\gamma (0)=26.6(1)$ mJ mol-Tm$^{-1}$K$^{-2}$
is obtained, indicating ${C_p \sim \gamma T}$ at the low temperaturea
limit. As shown in Fig.~\ref{fig3} (c), by applying magnetic field,
the zero temperature specific coefficient $\gamma (0)$ is
gradually suppressed. This means that part of $\gamma (0)$ could
be induced by the (quenched) disorders~\cite{Okamoto07,Kimchi}.
For magnetic field larger than 2 T where the disorder effect can be
reduced or removed~\cite{Kimchi}, $\gamma (0)$ reaches a finite value
of 3.7(5) mJ mol-Tm$^{-1}$ K$^{-2}$ for 2 T and 3(1) mJ mol-Tm$^{-1}$ K$^{-2}$
for 3 T, suggesting there is a constant density
of states at low energies and is consistent with a spin liquid state
with either spinon Fermi surfaces or a spinon quadratic node at the
spinon Fermi energy~\cite{LeeLeePRL,PhysRevLett.111.157203}.
Such a linear-$T$ heat capacity was also observed in the
organic QSL candidate $\kappa$-(BEDT-TTF)$_2$Cu$_2$(CN)$_3$~\cite{Yamashita08}
and EtMe$_3$Sb[Pd(dmit)$_2$]$_2$~\cite{Yamashita11}.
Thus, we propose the candidate spin liquid ground states to be a $\mathbb{Z}_2$
QSL with either spinon Fermi surfaces or a spinon quadratic touching node at the
spinon Fermi energy that give rise to a constant density of states
at low energies. A QSL with a continuous gauge group such as
$U(1)$ would have a larger density of states than a $\mathbb{Z}_2$ QSL
at low energies due to the soft gauge flucutations~\cite{Patrick1992,LeeLeePRL}.
When even larger magnetic fields are applied, a
field-induced gap is obtained. An activated form
${C_{p} \sim e^{-\Delta/T}}$ is then used to fit the 9 T data,
and we find an energy gap ${\Delta = 8.0(1)}$~K and a tiny residual
${\gamma (0) = 1.9(7) }$ mJ mol-Tm$^{-1}$ K$^{-2}$. In this high
field regime, the Tm$^{3+}$ local moments are polarized to the field
direction, the system is in an almost polarized state at low
temperatures (see Fig.~\ref{fig2} (c)), and the fitted gap
${\Delta = 8.0(1)}$ K is simply a magnon gap that is induced by
external magnetic fields. In Fig.~\ref{fig3} (b), we further subtract
the phonon contribution and the Schottky anomaly due to disorder effect
and show the calculated magnetic entropy up to 40 K. We find
the calculated entropy for zero field reaches $\sim$ R$\ln 2$
that is the entropy for spin-1/2 moments.

\subsection{ZF-$\mu$SR}

\begin{figure}[t]
\begin{center}
\includegraphics[width=8.5cm]{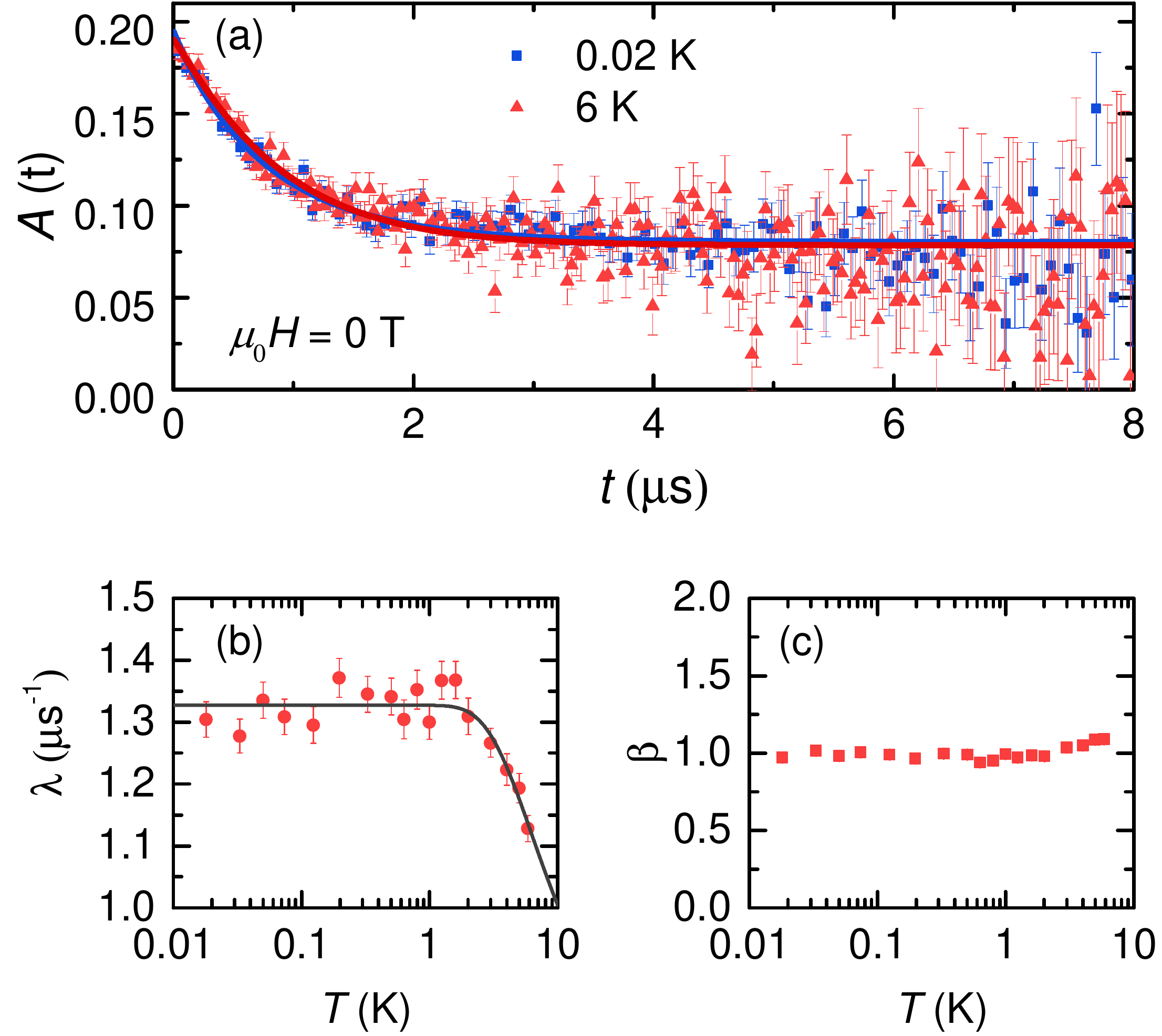}
  \caption{(a) The ZF-$\mu$SR spectra of Tm$_{3}$Sb$_{3}$Zn$_{2}$O$_{14}$
  measured at different temperatures. Lines are the fitting
  curves corresponding to the fitting function Eq.~(\ref{eq:fitting function}).
   The temperature dependence of fitting parameters $\lambda$ and $\beta$ are
   shown in (b) and (c), respectively. The gray curve in (b) is a guide
    to the eye, indicating the persistent spin dynamics below $\sim$2 K.}
  \label{fig4}
  \end{center}
\end{figure}

To further probe the magnetic properties, we implement the ZF-$\mu$SR
measurements on Tm$_{3}$Sb$_{3}$Zn$_{2}$O$_{14}$ down to 20 mK.
 The ZF-$\mu$SR spectra under two typical
temperatures were shown in Fig.~\ref{fig4}. No long
range order or spin freezing was detected in our sample. This
is evidenced by the absence of oscillation behavior (i.e. initial asymmetry loss)
in the spectra~\cite{Keren96,Zheng05,Miao16}. The static random
(time-reversal-breaking) field distribution was also excluded due
to the lacking of long time recovery to 1/3 of asymmetry,
suggesting that the relaxation was caused by dynamic
effects~\cite{Uemura85}. The ZF-$\mu$SR spectra were
further fitted by a single relaxation function:
\begin{equation}
\label{eq:fitting function}
A(t) = A_0+A_s(\frac{1}{3}+\frac{2}{3}e^{-(\lambda t)^{\beta}}),
\end{equation}
where $A_0$ is the constant background signal from silver sample holder,
$\lambda$ is the muon spin relaxation rate and $\beta$ is the stretched
exponent. Multiple components in the fitting function, originated from
multiple muon stopping sites, do not give better fit quality.
In a polycrystal sample, the local-field was randomly oriented
and the $\frac{1}{3}$ ($\frac{2}{3}$) term in the fitting function stands for
the local-field component parallel (perpendicular) to the initial muon spin.
A stretched exponential decay function was usually used
to describe the frustrated system\cite{Uemura94,Li16}.
In Fig.~\ref{fig4}(b), we show the temperature dependence of
relaxation rate $\lambda$. As the temperature is decreased, $\lambda$
gradually increases, saturates below 2 K, and remains almost constant
down to the lowest measured temperature 20~mK. The low temperature
plateau of $\lambda$ indicates the persistent spin dynamics and large
density of states at low energies~\cite{Uemura94}. This is consistent
with the large density of states from a QSL state with spinon Fermi
surface or a spinon quadratic touching node, instead of a gapped state.
Moreover, as is shown in Fig.~\ref{fig4} (c), the observed stretched exponent
$\beta$ is found to be $\sim$1 and is almost temperature independence,
suggesting the absence of obvious disorder/impurity induced relaxation
process~\cite{Yaouanc11}. We want to emphasize that previous study shows
that for some materials with non-Kramers ions, the measured $\mu$SR
response is dominated by an effect resulting from the muon-induced
local distortion rather than the intrinsic behavior
of the host compound, such as Pr$_2$\emph{B}$_2$O$_7$
(\emph{B} = Sn, Zr, Hf)~\cite{Foronda15}. However, it does not meet our
situation since no such static distribution of magnetic moments observed
in Pr$_2$\emph{B}$_2$O$_7$~\cite{Foronda15} was found in our experiments.
Further density functional theory calculations may give more information
of the effect of the induced muons. We argue that the dynamic relaxation is
caused by the Tm$^{3+}$ ions considering the larger effective magnetic moment.
Our result here is quite different from the behaviors in other frustrated
antiferromagnets such as spin glass in Y$_2$Mo$_2$O$_7$
and Tb$_2$Mo$_2$O$_7$~\cite{Dunsiger96} that show spin freezing,
or the magnetic ordered state in Gd$_2$Ti$_2$O$_7$~\cite{Dunsiger06}.


\section{DISCUSSION}

Here we explain the microscopic origin and discuss the candidate ground states
for Tm$_{3}$Sb$_{3}$Zn$_{2}$O$_{14}$. Since Tm$_{3}$Sb$_{3}$Zn$_{2}$O$_{14}$ is
a derived compound from the rare-earth pyrochlore material, we start from the
rare-earth ions on the pyrochlore lattice. Unlike the organics,
Tm$_{3}$Sb$_{3}$Zn$_{2}$O$_{14}$ is in the strong Mott regime with
well-localized $4f$ electrons. For an integer spin moment like the Tm$^{3+}$ ion,
the D$_{3d}$ crystal electric field of the pyrochlore lattice can create doubly
degenerate ground states that are called non-Kramers doublet, and
the transverse (longitudinal) component is the quadrupolar (dipolar)
component of the local moment~\cite{PhysRevB.78.094418,PhysRevLett.105.047201}.
As we have explained earlier, in a kagom\'{e} lattice, the lattice
symmetry does not protect the two-fold degeneracy of the non-Kramers
doublet any more. The small splitting between two states of the
non-Kramers doublet can then be modelled by a {\sl transverse} field
that acts on the quadrupolar component of
the local spin-1/2 moment~\cite{Savary17,PhysRevLett.118.107206}.
This should be distinguished from the Zeeman coupling to the longitudinal
(dipolar) component when the external field is applied to the
system~\cite{PhysRevB.96.195127,Yao17}.
Considering the spin-orbital-entangled nature of the Tm$^{3+}$
non-Kramers doublet, we propose that the magnetic properties of
Tm$_{3}$Sb$_{3}$Zn$_{2}$O$_{14}$ should be governed by the anisotropic
spin exchange interaction~\cite{PhysRevB.78.094418} and the transverse
Zeeman couplings~\cite{Savary17} on the kagom\'{e} lattice.
Thus the physics here is fundamentally different
from the Cu-based herbertsmithite material that is a well-known
kagom\'{e} lattice spin liquid candidate~\cite{Helton07,YLeeNature2012,
FuNMRScience2015,PhysRevLett.98.117205}.
The theoretical problem for future work is whether the proposed model
can give rise to the observed phenomena.

The possible QSL ground state, that we propose for Tm$_{3}$Sb$_{3}$Zn$_{2}$O$_{14}$,
is a $\mathbb{Z}_2$ QSL with either spinon Fermi surfaces or
a spinon quadratic band touching node at the Fermi level.
Our thermodynamic measurements cannot distinguish these two QSL states.
The inelastic neutron scattering measurement, however, should
observe different results for these two QSL states even with
polycrystal samples. For spinon Fermi surfaces, the low
energy scattering would extend a finite range of momenta
in the reciprocal space~\cite{Shen16,PhysRevB.96.054445}. In contrast,
the spinon quadratic band touching would produce the spinon continuum near the
Brillouin zone center at low energies. This distinction
does not require the angular information of the momenta.

Finally, we comment on the non-QSL possibility caused by the disorder effect in
Tm$_{3}$Sb$_{3}$Zn$_{2}$O$_{14}$.
Disorder effects are found in this material from XRD refinement.
Tm/Zn site disorder in kagom\'{e} lattice may modify the spin-spin
correlations, and tune the low temperature  ground state. The
disorder effect in the Cu-based kagom\'{e} structure materials
is still under debate~\cite{Han16,Feng17}. Recent theoretical
works show that disorder effect could lead to mimicry of spin
liquid like behavior in YbMgGaO$_4$~\cite{Zhu17}.
Unlike the Yb$^{3+}$ Kramers doublet in YbMgGaO$_4$ whose degeneracy is protected
by time reversal symmetry~\cite{YueshengPRL,YaodongPRB,Shen16,Martin17},
the non-Kramers doublet is more
susceptible to crystalline disorder. If there exists a crystal
field randomness for the Tm$^{3+}$ ion, it would lead to a
distribution of the transverse fields on the quadrupolar
component of the Tm$^{3+}$ non-Kramers doublet,
and the system can then be thought as a localized two-level
system with a random distribution of barrier heights
and asymmetry energies and could give a constant density of
low-energy states~\cite{Anderson1972,PhysRevB.49.12703}.
It is interesting both theoretically and experimentally
how to distinguish this scenario from the QSL proposal. Recently, Kimchi {\it et al.} analyzed the effect of quenched disorder on frustrated quantum magnets, and gave a phenomenological
description of the observed power-law magnetic specific heat with an anomalous exponent~\cite{Kimchi18}. In their random-singlet-inspired picture, the anomalous $C(T)$ exponent is an effective exponent over a range of temperature scales, and can take different values depending on the disorder distribution~\cite{Kimchi18}. Therefore, we can not rule out that the exotic properties are induced by the Tm/Zn site-mixing disorder in Tm$_{3}$Sb$_{3}$Zn$_{2}$O$_{14}$. On the other hand, we also noticed that disorder can induce QSL state in spin ice pyrochlores~\cite{Savary17}.

\section{CONCLUSION}

To summarize, we have studied the crystal structure, magnetic properties,
specific heat and $\mu$SR spectra of Tm$_{3}$Sb$_{3}$Zn$_{2}$O$_{14}$. No
long range magnetic order was observed down to 20 mK. Spin freezing behavior
was precluded by A.C. susceptibility measurements. A linear-$T$
dependence of low temperature magnetic specific heat was observed,
indicating a constant density of low-energy states.  The
ZF-$\mu$SR results suggest persistent spin dynamics at low
temperature range. We propose possible QSL ground states
and provide the microscopic origin for the magnetic properties
of Tm$_{3}$Sb$_{3}$Zn$_{2}$O$_{14}$.

\emph{Acknowledgments.}---We wish to thank B. Hitti, and D. Arsenau
of the TRIUMF CMMS group for the assistance during the experiments.
We also acknowledge Martin Mourigal for an email correspondence,
Yi Zhou from Zhejiang University for a discussion, Patrick Lee
for a recent comment about the specific heat. We thank Zhong Wang from
IAS Tsinghua Univeristy for hospitality where this paper is completed.
This work is supported by the National Key Research and Development
Program of China (Nos.2016YFA0300503 (L.S.), 2016YFA0301001 (G.C.),
the start-up fund and the first-class university construction
fund of Fudan University (G.C.), the thousand-youth-talent program of
China (G.C.), and the National Natural Science Foundation of China
No.11474060 (L.S.) and No.11774061 (L.S.).

%

\end{document}